\pdfoutput=1
\documentclass[12pt,a4paper,twocolumn]{article}

\usepackage[english]{babel}
\usepackage[utf8]{inputenc}
\usepackage[T1]{fontenc}
\usepackage{geometry}
\usepackage{setspace}
\usepackage[scaled]{helvet}
\usepackage{graphicx}

\usepackage{titlesec}

\titleformat{\section}
  {\normalsize\bfseries}
  {}
  {0pt}
  {}
  [\titlerule]

\usepackage[hidelinks]{hyperref}
\title{Beyond the GPU: The Strategic Role of FPGAs in the Next Wave of AI}
\author{Arturo Urías Jiménez}
\date{Instituto Tecnológico y de Estudios Superiores de Monterrey}

\begin{document}
\onehalfspacing

\maketitle

\section{Introduction}

AI acceleration has been dominated by GPUs, but the growing need for lower latency, energy efficiency, and fine-grained hardware control exposes the limits of fixed architectures. In this context, Field-Programmable Gate Arrays (FPGAs) emerge as a reconfigurable platform that allows mapping AI algorithms directly into device logic. Their ability to implement parallel pipelines for convolutions, attention mechanisms, and post-processing with deterministic timing and reduced power consumption makes them a strategic option for workloads that demand predictable performance and deep customization.

Unlike CPUs and GPUs, whose architecture is immutable, an FPGA can be reconfigured in the field to adapt its physical structure to a specific model, integrate as a SoC with embedded processors, and run inference near the sensor without sending raw data to the cloud. This reduces latency and required bandwidth, improves privacy, and frees GPUs from specialized tasks in data centers. Partial reconfiguration and compilation flows from AI frameworks are shortening the path from prototype to deployment, enabling hardware--algorithm co-design.

\section{What is an FPGA?}

Unlike most chips, whose logical functions are fixed once manufactured, FPGAs offer unprecedented flexibility by allowing their internal logic to be configured and reconfigured to suit various applications even after production; this essential feature makes them a highly versatile tool in computing \cite{noauthor_fpgas_nodate}.

Internally, an FPGA is composed of six main elements: programmable input/output units (IO), programmable logic units (CLB), routing resources, integrated block memory (BRAM), complete clock management and dedicated hardware modules, and underlying functional units \cite{boudjadar_dynamic_2025}. Logic units are implemented using look-up tables (LUTs), often combined with other components such as flip-flops to form independent logic blocks; most FPGA functions are performed through these logic blocks \cite{noauthor_fpgas_nodate}.

FPGAs offer a unique combination of speed, programmability, and flexibility that delivers performance without the complexity and cost of developing custom chips (ASICs). This adaptability, inherent to their reconfigurable architecture, makes them particularly suitable for a range of Machine Learning (ML) applications, from edge computing to the demands of large-scale data centers \cite{noauthor_fpgas_nodate}.

\section{FPGAs and AI Independence from Servers}

FPGAs are crucial to decoupling AI from servers in IoT applications and to increasing security and privacy, thanks to their ability to enable edge computing and embedded AI \cite{li_dataflow_2025} \cite{boudjadar_dynamic_2025}.

ML solutions---especially CNN-based ones---have historically relied on the cloud, where the actual network is deployed on powerful compute servers. However, for many application domains, such as control systems and IoT, this dependency is shifting significantly toward embedded computing.

FPGAs support this independence by enabling \textit{pervasive computing} that makes computation ubiquitous and reduces reliance on connectivity. With intelligence on the device, decision-making and inference can occur without a constant connection to the cloud or a central server \cite{boudjadar_dynamic_2025}.

They also reduce communication costs, since they allow processing and feature extraction directly at the data source. This is vital for IoT, where continuously transmitting large volumes of raw data to a central server is costly and inefficient. At the same time, they offer greater security and privacy because they avoid unnecessary exposure of sensitive data during transmission; as data never leaves the local network or device, interception or compromise risk decreases \cite{boudjadar_dynamic_2025}.

\section{FPGA vs GPU}

To achieve secure, local processing, we need hardware platforms capable of accelerating AI on the device itself. The most widely used include FPGAs and GPUs. Understanding their differences helps choose the most suitable option based on system needs, costs, and design goals. Both are very powerful, but their distinct architectures make them better suited to different tasks \cite{noauthor_fpgas_nodate} \cite{boudjadar_dynamic_2025}.

GPUs are a type of specialized circuitry designed to quickly manipulate memory and accelerate image creation. They are built for high throughput and excel at parallel processing tasks, making them a preferred choice for the thousands of simultaneous operations required for training and inference of neural networks \cite{noauthor_fpga_2024}.

GPUs are favored for their prowess in parallel processing in compute-intensive ML applications. Powerful graphics processors handle demanding tasks such as high-performance computing (HPC) and are especially capable of computing large numbers of matrix multiplications concurrently, significantly accelerating training times for large deep learning models. Generally, GPUs are preferred for heavier tasks such as training and running large, complex models, or in data center environments \cite{yan_survey_2024}.

Energy efficiency is an area where FPGAs have a clear advantage over GPUs. The impressive processing power of GPUs comes at the cost of energy efficiency and high power consumption. This high consumption can increase operational expenses and impact environmental concerns. In contrast, FPGAs use less power than other processors. Implementing AI on FPGAs helps optimize energy efficiency and, in data center environments, their high efficiency helps mitigate cooling costs, supporting the development of ``greener'' AI technologies \cite{noauthor_fpgas_nodate}.

In terms of flexibility and latency, FPGAs outperform GPUs. GPUs are far less flexible than FPGAs, offering fewer opportunities for optimization or customization for specific tasks, as their architecture is fixed. By contrast, programmability is central to FPGA design, enabling fine-tuning and prototyping---useful in the emerging field of deep learning. FPGAs have flexible, customizable I/O and can guarantee low and deterministic latency, crucial for real-time applications such as radar signal processing, autonomous vehicles, and telecommunications. Although FPGAs are valued for energy efficiency, their inherently lower power makes them less suitable for the most demanding tasks \cite{xu_fpga_2023}.

Finally, regarding programming and ecosystem, GPUs enjoy strong vendor support and robust frameworks, including CUDA and OpenCL. On the other hand, FPGA programming is a well-known challenge, as hardware requires specialized expertise \cite{xu_fpga_2023}. Programming and reprogramming FPGAs is laborious and time-consuming, potentially delaying deployments. However, software-based programming models and high-level tools are increasingly used to reduce the expertise required \cite{noauthor_fpgas_nodate}\cite{li_edge_2024}.

\section{Implementation Examples}

To support this comparison, consider the following examples:
The article \textit{An Edge AI System Based on FPGA Platform for Railway Fault Detection} \cite{li_edge_2024} describes an edge AI system for railway inspection that uses a Xilinx ZCU104 FPGA to run, in real time, a lightweight CNN that detects defects on tracks from camera images. This platform replaces manual inspection, achieves 88.9\% accuracy, processes 60 images per second with latencies below 17 ms, and transmits results via an ESP8266 module to a graphical interface. Its energy efficiency surpasses a GPU by 1.39$\times$ and a CPU by 4.67$\times$, demonstrating that FPGAs are an ideal solution for implementing high-performance, low-power edge AI \cite{li_edge_2024}.

\begin{figure}[h]
\centering
\includegraphics[width=0.45\textwidth]{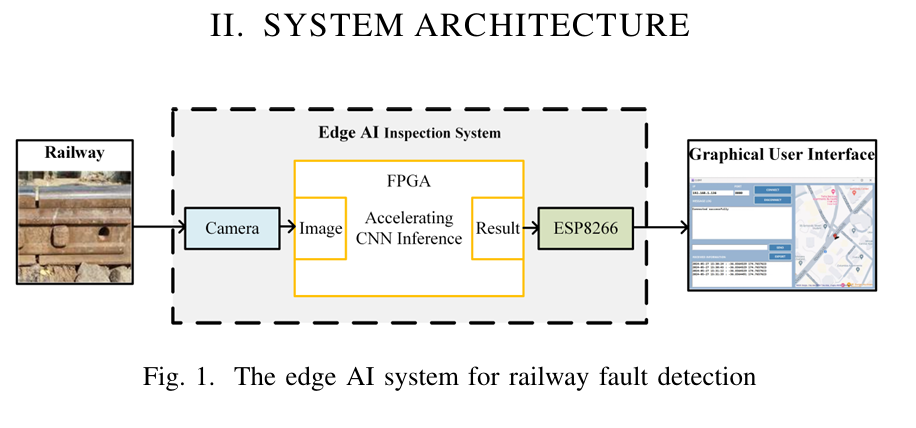}
\caption{Diagram of the proposed system. \cite{li_edge_2024}.}
\label{fig:sistema}
\end{figure}

Microsoft's Project Brainwave aims to meet low-latency and high-bandwidth demands in AI applications without sacrificing accuracy or quality. It implements and optimizes neural cores (softcore NPU) and instructions directly on FPGAs, automatically partitions trained deep networks into sub-regions, and executes them on a joint FPGA--CPU architecture with a full acceleration flow for DNNs. The system allows users to deploy models and obtain hardware acceleration without designing hardware, maintaining real-time, low-cost services. In tests on Bing, Brainwave achieved more than a tenfold latency reduction and greater model capacity compared to CPUs \cite{xu_fpga_2023}.

\begin{figure}[h]
\centering
\includegraphics[width=0.45\textwidth]{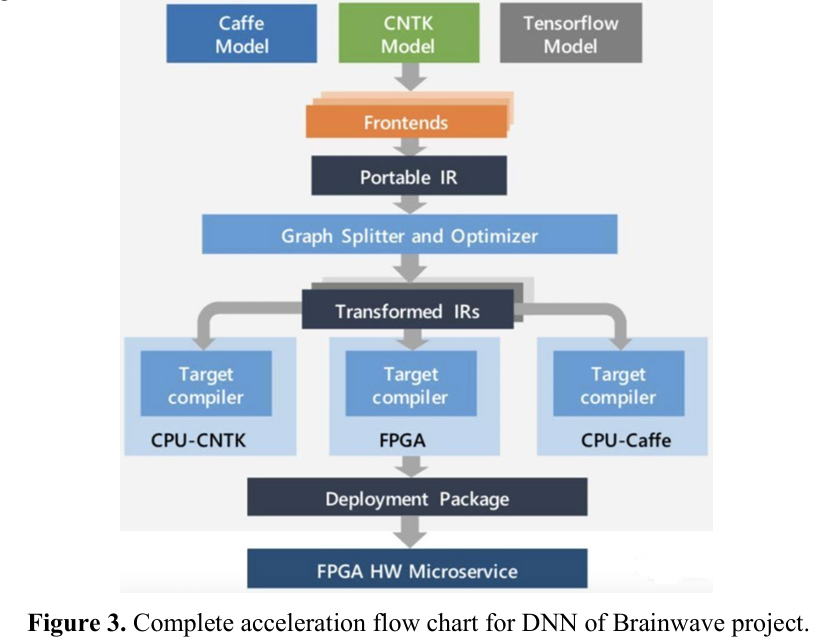}
\caption{Diagram of the proposed system. \cite{xu_fpga_2023}.}
\label{fig:sistema}
\end{figure}
\newpage
\newpage

\section{Conclusion}

From these examples we can confirm the advantages in energy, efficiency, and sustainability, positioning FPGAs as a crucial and indispensable component in the future AI ecosystem. This reconfigurable technology is intrinsically superior in energy efficiency compared with CPUs and GPUs, which is vital not only to reduce operational costs but also to drive the development of more sustainable, ``green'' AI. FPGAs achieve this superiority through deep hardware customization tailored to fixed inference tasks, eliminating unnecessary energy consumption by precisely adapting the architecture to application requirements. This efficiency is reinforced by the ability to use low-precision computation (quantization to reduced bit-widths), minimizing computational complexity and memory demands without sacrificing model accuracy, as well as advanced memory and dataflow optimization strategies that reduce data movement and access to power-hungry off-chip memories \cite{noauthor_fpgas_nodate}.\newline\newline\newline\newline\newline\newline
\newline\newline\newline\newline

\end{document}